\documentclass{emulateapj}

\newcommand{\etal}{{et al.}}
\newcommand{\eg}{{\it e.g.,}}
\newcommand{\ie}{{\it i.e.,}}
\newcommand{\kms}{km~s$^{-1}$~}
\newcommand{\ea}{4C\,23.56\,ER1}
\newcommand{\eb}{4C\,05.84\,ER1}

\begin{document}

\title{Morphologies of Two Massive Old Galaxies at $z\sim2.5$\altaffilmark{1}}

\author{Alan Stockton\altaffilmark{2} and Elizabeth McGrath}
\affil{Institute for Astronomy, University of Hawaii, 2680 Woodlawn
 Drive, Honolulu, HI 96822}

\author{Gabriela Canalizo}
\affil{Department of Physics and Astronomy, University of California, Riverside, CA 95521}

\author{Masanori Iye}
\affil{Optical and Infrared Astronomy Division, National Astronomical Observatory of Japan, Mitaka, Tokyo, 181-8588, Japan}

\and

\author{Toshinori Maihara}
\affil{Department of Astronomy, Kyoto University, Kitashirakawa-Oiwake-cho, Sakyo-ku, Kyoto 606-8502, Japan}

\altaffiltext{1}{Based in part on observations made with the NASA/ESA Hubble
Space Telescope, obtained at the Space Telescope
Science Institute, which is operated by the Association of Universities
for Research in Astronomy, Inc., under NASA contract NAS 5-26555. These
observations are associated with program no.~10418. Additional data were collected 
at the Subaru Telescope, which is operated by the National Astronomical Observatory 
of Japan, and at the W. M. Keck Observatory, which is operated as a scientific partnership
among the California Institute of Technology, the University of California, and the
National Aeronautics and Space Administration.  The Observatory was made possible
by the financial support of the W. M. Keck Foundation.}

\altaffiltext{2}{Also at Cerro Tololo Inter-American Observatory, Casilla 603, La Serena, Chile.}

\begin{abstract}
We present the results of NICMOS imaging of two massive galaxies photometrically selected to have old
stellar populations at $z\sim2.5$.  Both galaxies are dominated by apparent 
disks of old stars, although one of them also has a small bulge
comprising about 1/3 of the light at rest-frame 4800 \AA.  The presence
of massive disks of old stars at high redshift means that at least some massive
galaxies in the early universe have formed directly from the dissipative 
collapse of a large mass of gas. The stars formed in disks like these may have
made significant contributions to the stellar populations of  massive spheroids at
the present epoch.
\end{abstract}


\section{Introduction}
Considerable observational evidence has built up over the past few years that
a substantial
fraction of the massive galaxies around us today were already massive at very early epochs.
This evidence comes primarily from three sources:
\begin{itemize}
\item  Studies of local massive elliptical galaxies indicate that the stars in the most
massive galaxies generally formed very early and over very short time intervals
\citep{pee02,tho05,nel05,ren06}.  Stars
in less massive spheroids formed, on average, later and over longer time spans.
\item Massive galaxies in clusters show little evidence for significant evolution up to at
least redshift $\sim1$ \citep[\eg][]{deP07,sca07}.
\item Direct observations of massive galaxies at redshifts $\gtrsim1.5$ that are dominated
by already old stellar populations show that significant numbers of massive galaxies
were in place at even earlier epochs \citep*[\eg][]{sto04,mcC04,vanD04,lab05,dad05,
red06,pap06,kri06,abr07}.
\end{itemize}

Although the existence of massive galaxies at high redshifts is now well documented, there have
been only a few high-resolution studies of their morphologies (e.g., \citealt{yan03,sto04,zir07,tof07}).  
Morphologies are important, because they may well 
retain signs of the formation history of the galaxies.  This is particularly true for galaxies
that show little or no recent star formation, so that we are able to observe relatively clean
examples of the stellar population that formed earliest and that comprises the bulk of the 
mass of the galaxy.
In this paper, we present deep {\it Hubble Space Telescope} ({\it HST}) NICMOS imaging 
of two galaxies with virtually
pure old stellar populations at $z\sim2.5$.  In \S~2, we briefly recount how these galaxies were
selected.  In \S~3, we describe the observations and reduction procedures.  In \S~4 and
\S~5, we
analyze model fits to the images to determine morphologies, and in \S~6 we discuss the implications
of our conclusions.  We assume a flat cosmology with $H_0 = 73$ \kms\ Mpc$^{-1}$ and
$\Omega_M = 0.28$.

\section{Identifying Galaxies with Old Stellar Populations at High Redshifts}

Our procedure for selecting galaxies with old stellar populations is described in some detail
in \citet{sto04}; here we give a brief synopsis.  We observe fields of radio sources in certain
specific redshift ranges, selecting galaxies with photometric redshifts consistent with that
of the radio source.  Radio sources generally serve as beacons for some of the more
overdense regions in the early universe.  Furthermore, the specific redshift ranges 
selected are chosen to optimize discrimination with standard filter passbands 
between old stellar populations and highly reddened star-forming galaxies. One of these redshift
ranges is $2.3<z<2.7$, for which the 4000 \AA\ break, strong in old stellar populations,
falls between the $J$ and $H$ bands.  We have used the \citet{bru03} (BC03) spectral synthesis 
models, and, more recently, preliminary versions of the \citet{cha07} (CB07) models, to
evaluate and optimize our photometric selection of old stellar populations at various
redshifts.  The preliminary CB07 models include more realistic prescriptions for thermally pulsing 
asymptotic-giant-branch stars (\citealt{mar07}; see also \citealt{mar05}).  Although at low redshifts
(and for some SEDs at high redshifts) the new models can significantly lower the masses estimated 
from $K$-band photometry, at the redshifts we are considering here for nearly pure old stellar
populations, the masses (and ages) change hardly at all.  The main effect of using the newer
models is to reduce the amount of reddening required to obtain a good fit.

If a stellar population were to have an age of
2 Gyr at $z=2.5$ (corresponding to all of the stars forming at $z=9$), its observed colors 
would be $J\!-\!K\approx3.0$ and $J\!-\!H\approx2.1$.  We use a photometric sieve procedure to optimize
the selection with respect to available observing time, first obtaining relatively short $J$ and
$K'$ integrations (typically 5 $\sigma$ at $J=23$ and $>10$ $\sigma$ at $K'=20$).  If any objects
with $J\!-\!K'\sim3$ are found, we then obtain $H$ and deeper $J$ imaging.  Finally, for fields
with objects matching the expected spectral-energy distributions of an old stellar population
at the redshift of the radio source, we attempt to obtain deep imaging at shorter wavelengths
(usually either $R$ or $I$) to set constraints on any residual star formation.

Among the galaxies found by this technique are one each in the fields of the radio galaxy
4C\,23.56 \citep{sto04} and the quasar 4C\,05.84.  
We refer to these galaxies as \ea\ and \eb; they are both luminous objects,
and they have stellar populations that appear to be overwhelmingly dominated by old stars.

\section{Observations and Data Reduction}

\subsection{Ground-Based Optical and Near-IR Observations}

We obtained most of the near-IR observations ($J$, $H$, and $K'$) with the CISCO IR camera
\citep{mot02} on the 8.2 m Subaru Telescope \citep{iye04} in observing runs on 
2000 November 8 (UT),
2001 August 5 and 6, and 2002 May 30--June 1.  The images have a scale of
0\farcs105 pixel$^{-1}$ and a field of $\sim1\farcm8$. In addition, we carried out deep $R$-band
imaging of both fields with the Echelle Spectrograph and Imager (ESI;
\citealt{she02}) on the Keck II Telescope on 2002 August 7.  Both the IR and optical
imaging were reduced according to standard procedures using our own IRAF scripts.
The calibrations used observations of UKIRT Faint Standards \citep{haw01,leg06}
for the IR photometry and Landolt fields \citep{lan92} for the $R$-band imaging. 

We also observed \ea\ at $K'$ with the Subaru 36-element
curvature-sensing adaptive optics (AO) system \citep{tak04} 
and the Infrared Camera and Spectrograph
(IRCS; \citealt{kob00}) on 2002 August 17.  These results were reported by
\citet{sto04}, but we will refer to them again in this paper.  We used IRCS without
the AO system, but with excellent natural seeing (final images have FWHM of 0\farcs35)
to obtain a very deep image of the 4C\,05.84 field in the $K$ filter on 2004 August 1.
Finally, we obtained $J$-band imaging of the 4C\,05.84 field with NIRC2 and the Keck II
laser-guide-star adaptive-optics system on 2007 August 21.

\subsection{{\itshape Hubble Space Telescope} NICMOS Observations}

The NICMOS observations used the NIC2 camera (0\farcs075 pixel$^{-1}$) and the
F110W and F160W filters.  They were obtained on UT 2005 January 3 (\eb, F160W, 
total exposure 5376 s), 2005 January 4 (\ea, F160W, total exposure 8192 s), 
2005 January 8 (\eb, F110W, total exposure 8448 s), and 2005 May 16
(\ea, F110W, total exposure 11264 s) as part of {\it HST} program 10418. 
After doing a first-pass combination
of the images to get a rough idea of the quality of the data, we went back to the
{\it calnica} processed images and corrected these for bias offsets and inverse
flatfield effects using the STSDAS {\it pedsky} task.  Most of the F110W images
were obtained in orbits impacted by passages through the South Atlantic Anomaly
(SAA) and needed special processing.  We used the IDL routine saa\_clean.pro
\citep{ber03} to generate an image of the persistence from the routinely taken
post-SAA dark images and subtract it from the science images.  Finally,
for these images, {\it pedsky} was run again to remove any residual bias
pedestals.

We then generated a bad-pixel
mask from the data quality file, adding an additional mask for the coronographic
occulter, which produces background that is detected in both filters. Most of
the cosmic rays were removed with the contributed IRAF procedure {\it lacos\_im}
\citep{vanD01}.  At this point the images from the individual dither positions were
combined onto a subsampled grid with the STSDAS {\it drizzle} task 
to produce the final image. To choose the optimum {\it drizzle} parameters for
our purposes, we performed a series of tests with artificial PSFs generated by
Tiny Tim\footnotemark[3]. 
\footnotetext[3]{Tiny Tim was written by J. Krist and can be
found at http://www.stsci.edu/software/tinytim/tinytim.html}
We ended up choosing a drop size of 0.7 and a
subsampling factor of 2. The final combined images had FWHM of 0\farcs133 
for the F160W images and 0\farcs115 for the \ea\ F110W image.
We were unfortunately unable to produce a useful image
of \eb\ in the F110W band because of a combination of the object's low surface
brightness at that wavelength and residual effects on the detector of the 
previous SAA passage.  The final $3 \sigma$ surface brightness limits 
(in the Vega system) were $\mu_{160} \approx 22.3$ and $\mu_{110} \approx 23.3$
for \ea, and $\mu_{160} \approx 22.0$ for \eb.

The drizzling process inevitably introduces some level of correlation between adjacent
pixels.  Where we needed to estimate absolute errors (such as for our radial-surface-brightness
plots), we made a statistical correction to the error determinations, following the prescriptions
of \citet{fru02}

Although we obtained images of stars for point-spread-function (PSF) determination
at the ends of some of the orbits, PSFs modeled from TinyTim were
quite consistent with the stellar profiles.  Subtracting TinyTim models
from the observed stars gave residuals that were less than 1.5\%\ (rms) of the peak over
the central FWHM region (and much lower outside this region), with maximum pixel 
deviations of 3\%.  The TinyTim profiles also have the advantage that they can be 
generated on a subsampled grid to minimize interpolation errors in matching the
profiles to the undersampled NICMOS2 images.  We accordingly used subsampled 
TinyTim model profiles in our analysis.

\section{4C\,23.56\,ER1}

In the field of the $z=2.483$ radio galaxy 4C\,23.56, our photometric selection procedure picked
out a galaxy that had previously been noted as a very red object by \citet{kno97}.
As mentioned above, we have already reported on our Subaru AO/IRCS imaging of \ea\
(\citealt{sto04}; there, the galaxy is referred to as 4C\,23.56\,KC68). The main conclusions 
of that paper were that (1) the galaxy indeed has a redshift close to that of the radio galaxy,
(2) the best fit to the photometry is an old (2--3 Gyr) stellar population with little reddening
($\lesssim0.2 A_V$ of \citealt{cal00} extinction), and that (3) the morphology of this massive,
old galaxy looked surprisingly disklike, with a projected axial ratio of 0.3 and a S\'{e}rsic 
index of 1.5.  Because the details of the previous observations are available in that paper,
and in a brief follow-up report \citep{sto07},
we restrict our discussion here to a comparison of the NICMOS imaging with the 
previous AO imaging.

\begin{figure*}[!bt]
\epsscale{1.0}
\plotone{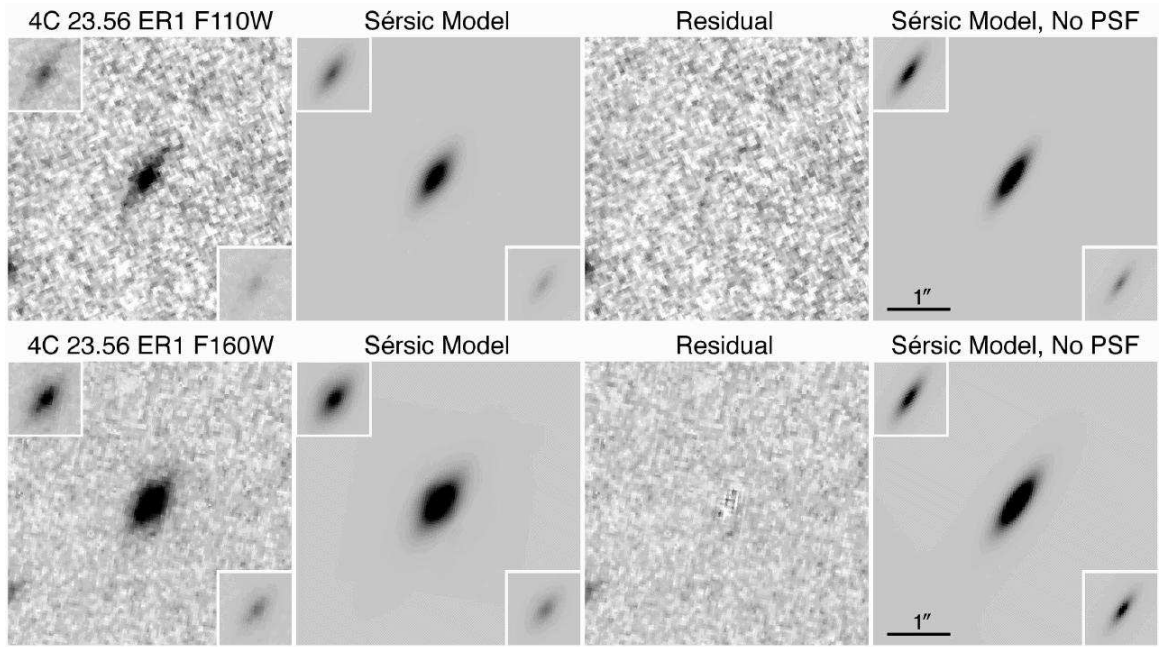}
\caption{NICMOS2 images of 4C\,23.56\,ER1 in the F110W and F160W filters.
The best-fit {\sc galfit} S\'{e}rsic models, convolved with the PSF, are shown in 
the second panel of each 
row, the difference between the observed images and the models in the third
panel, and the models without convolution with the PSF in the last panel.
Insets show lower-contrast versions of the images.  North is up and East to the
left for this and all following images.}\label{4c23mos}
\end{figure*}

The NICMOS F110W and F160W images are shown in Fig.~\ref{4c23mos}, along with
the best-fit S\'{e}rsic models (convolved with the PSF), the residuals from the subtraction 
of the models from the data, and, again, the best-fit models (but {\em without} convolution
with the PSF).  We determine total magnitudes for \ea\ from the S\'{e}rsic models, finding
(on the Vega system) $m_{F110W} = 23.39 \pm 0.17$ and $m_{F160W} = 20.82 \pm 0.04$
after correction for Galactic reddening \citep{sch98}.  The quoted uncertainties include only
sky noise and uncertainty in the sky level; they do not include any deviations between the
models and the data (which are in any case quite small over the region of good S/N for the
data) or other potential systematic effects.

\begin{figure}[!bt]
\epsscale{1.0}
\plotone{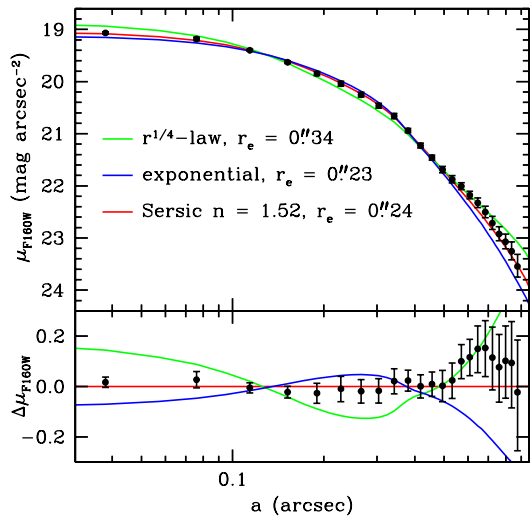}
\caption{Radial-surface-brightness profile of the NICMOS2 F160W image of
\ea, with best-fit $r^{1/4}$-law, exponential, and S\'{e}rsic profiles shown.
The upper panel shows the profiles, and the lower panel shows the
deviations of the observed profile and the two other models from the best-fit
S\'{e}rsic profile. Sample points in this and subsequent plots are at intervals
of 1 subsampled pixel in the drizzled images (0\farcs038) along the major
axis, so data values
and errors for adjacent points are fairly strongly correlated because of 
drizzling, PSF smearing, and compression of the scale along the minor
axis.}\label{4c23hrsb}
\end{figure}

\begin{figure}[!tb]
\epsscale{1.0}
\plotone{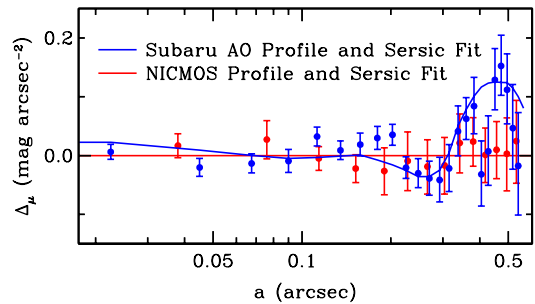}
\caption{Comparison of normalized differential surface-brightness profiles for \ea\
in the region of overlap.  
All profiles are shown relative to the NICMOS2 F160W S\'{e}rsic fit, and the difference
between F160W and $K'$ magnitudes has been removed via a simple slide fit.
The red points and line show the F160W profile and S\'{e}rsic fit, respectively, 
and the blue points and line show the Subaru AO $K'$ profile and fit.}\label{4c23drsbp}
\end{figure}

We discuss the F160W image first, since it has a much higher S/N ratio than
does the F110W image.
Figure \ref{4c23hrsb} shows the radial-surface-brightness profile for 
the F160W image of \ea, along
with the best-fit $r^{1/4}$-law, exponential, and S\'{e}rsic profiles, determined
using {\sc galfit} \citep{pen02}.
Among these, the S\'{e}rsic profile clearly gives the best fit, as expected,
because of the extra degree of freedom in the model.
The S\'{e}rsic profile has an index $n = 1.52\pm0.06$, 
an effective radius $r_e = 0\farcs24\pm0.01$, and axial ratio $b/a = 0.32$.
The uncertainties in $n$ and $r_e$ have been estimated by re-running the
models with the sky level set $1\sigma$ above and below its median value.  From
our Subaru AO imaging in the $K'$ band \citep{sto04}, we had obtained
$n = 1.49$, $r_e = 0\farcs22$, $b/a = 0.33$, so the two independent profiles
in different bands are in remarkably good agreement.  Although the AO imaging had slightly 
better FWHM, and the two datasets had similar S/N near the center of the 
galaxy, the NICMOS2 data extends farther in semi-major axis because
of its lower sky background.  We show a comparison of the two profiles in the
region of overlap in Fig.~\ref{4c23drsbp}.
Both the $r^{1/4}$-law and exponential profiles fit the observed profile poorly.  Adding
a small ($r_e=0\farcs1$), weak (14\%\ of total light)  bulge component to an exponential 
profile with an $r_e = 0\farcs26$ gives a fit that is as good
as that of the S\'{e}rsic profile within a semi-major axis of 0\farcs6 but somewhat
worse beyond this radius.

\begin{figure}[!tb]
\epsscale{1.0}
\plotone{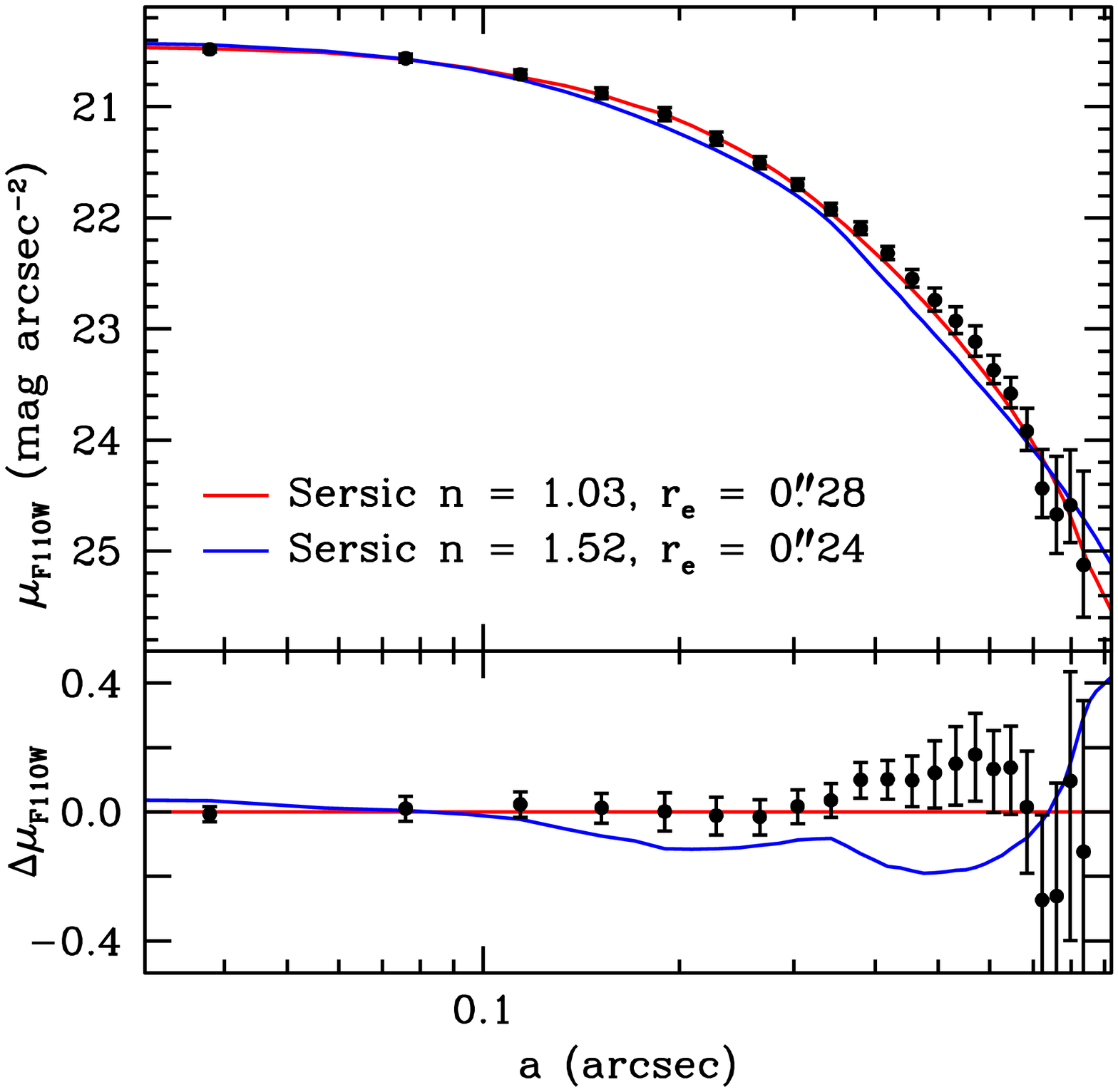}
\caption{Radial-surface-brightness profile of the NICMOS2 F110W image of
\ea, with best-fit S\'{e}rsic profile shown (red trace).  Also shown is the
best-fit F160W S\'{e}rsic profile from Fig.~\ref{4c23hrsb}, shifted by a
constant magnitude offset to match the F110W points (blue trace).
The upper panel shows the profiles, and the lower panel shows the
deviations of the observed F110W profile and the F160W model from the best-fit
S\'{e}rsic profile to the F110W data.}\label{4c23jrsb}
\end{figure}

The F110W image shown in Fig.~\ref{4c23mos}, which samples the morphology
shortward of the 4000 \AA\ break (assuming that \ea\ has the same redshift as
4C\,23.56 itself), superficially has an even more ``disky'' appearance
than does the F160W image.  This is partly due to the sharper PSF at this
wavelength:  notice that the best-fit S\'{e}rsic models without PSF convolution
look much more similar than do the models with PSF convolution.  Nevertheless,
there may be a detectable difference in morphology in the two bands.
The F110W S\'{e}rsic model has an index $n = 1.03\pm0.10$ (\ie\ essentially a pure
exponential), an effective radius of $0\farcs28\pm0\farcs02$, and $b/a = 0.31$.  The 
radial-surface-brightness profile is shown in Fig.~\ref{4c23jrsb}, along with
the best-fit S\'{e}rsic model and the F160W S\'{e}rsic model (adjusted by a constant
magnitude offset to approximately match the F110W points).
The observed differences are barely significant, given the uncertainties, but they seem 
to indicate a small color gradient, such that the outer parts of
the galaxy are slightly bluer (at least out to a semi-major axis of 0\farcs7, at which point
the uncertainty in the sky background level becomes dominant).  
This cannot be a large effect because of the tight upper limits
on the $R$ and $I$-band magnitudes (see \citealt{sto04}).  Nevertheless, it does suggest
a possible slight decrease in mean age and/or mean metallicity of the stellar population as one 
progresses from the center to the outskirts of the galaxy.

\section{4C\,05.84\,ER1\label{4c05}}

\begin{figure}[!tb]
\epsscale{1.0}
\plotone{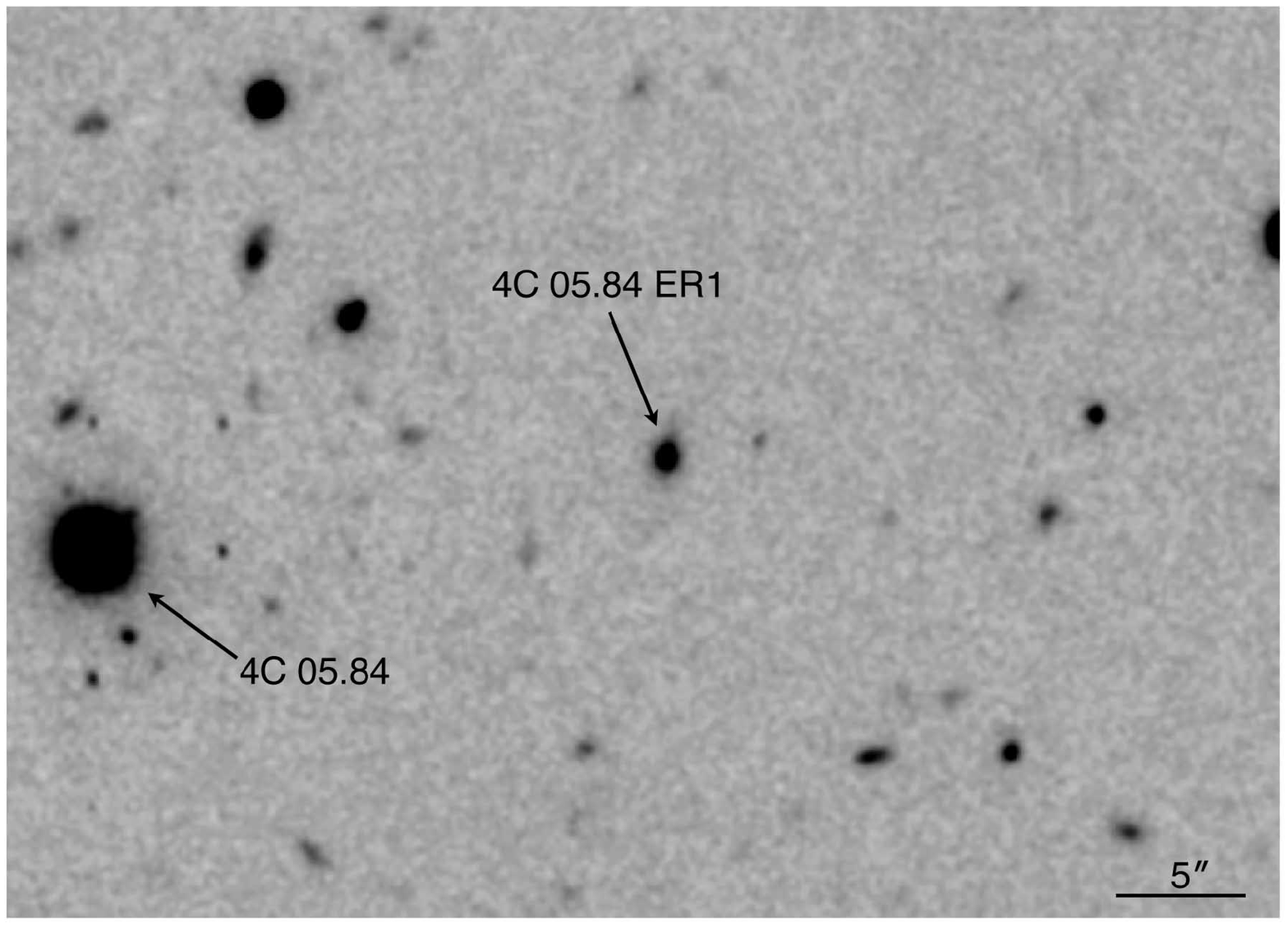}
\caption{The field of \eb.  The image is a deep (6750 s) $K$-band integration with
the Subaru Infrared Camera and Spectrograph in 0\farcs35 seeing. North is up
and East to the left.}\label{4c05field}
\end{figure}

\begin{figure}[!bt]
\epsscale{1.0}
\plotone{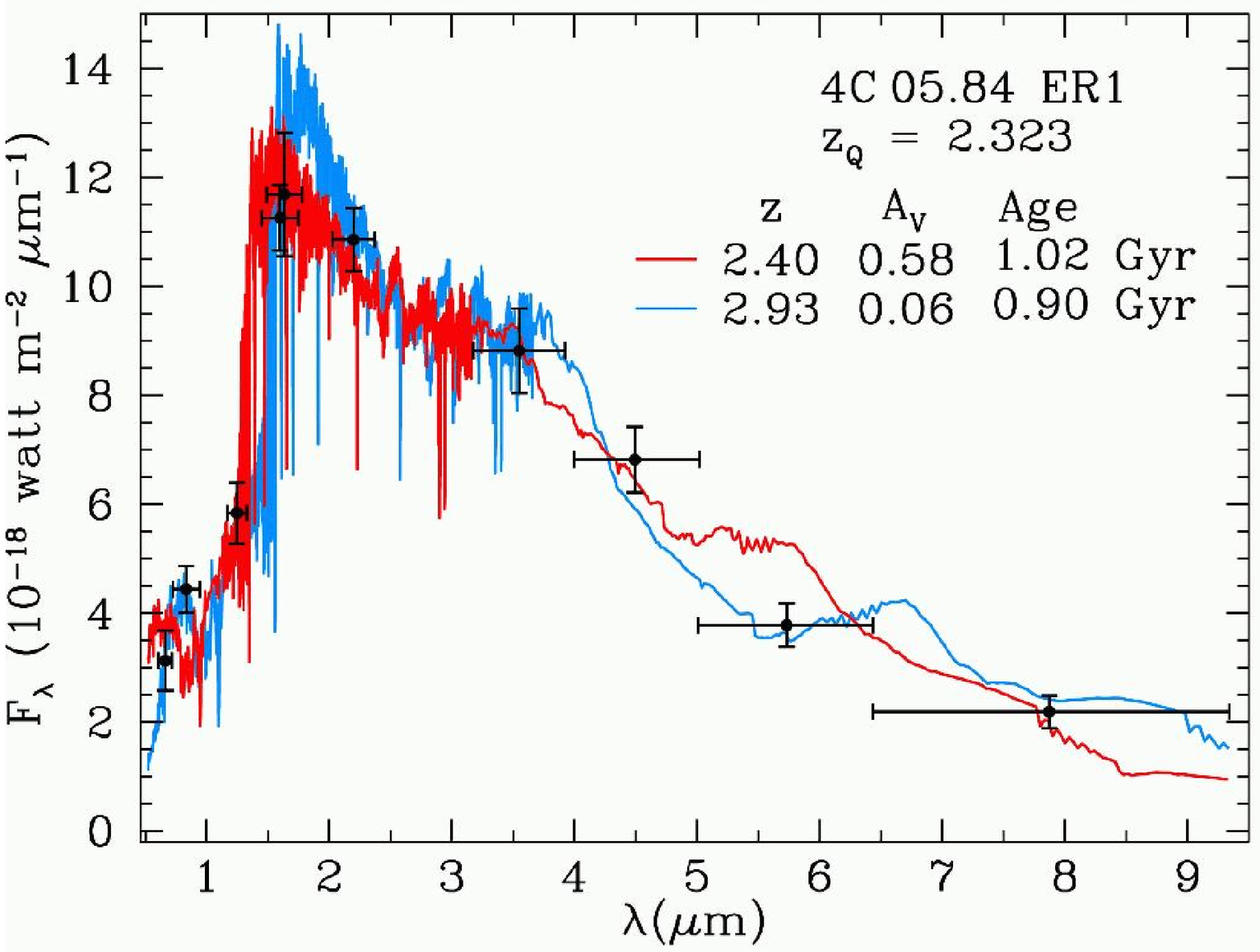}
\caption{Spectral-energy distribution for \eb.  The blue curve shows the best-fit CB07 
stellar population model to the photometry, at a redshift $z=2.93$.  The red curve shows
the best-fit model with a redshift close to that of 4C\,05.84 itself ($z_Q$).  The redshifts
$z$, Calzetti-law extinctions $A_V$, and ages of the stellar populations are indicated for
each model.  See the text for details.}\label{4c05sed}
\end{figure}

\eb\ was found in the field of the $z=2.323$ quasar 4C\,05.84 (Fig.~\ref{4c05field}).
The SED of \eb\ is shown in Fig.~\ref{4c05sed}, including photometry from 
our Spitzer IRAC images, which will be discussed elsewhere in more detail in 
the context of a larger sample of objects.  While, for \ea, only upper limits
at $R$ and $I$ bands have been obtained, for \eb\ we have detections at $R=24.6$
and $I=23.4$, indicating the presence of some younger stars. 
In attempting to fit the observed SED, we have explored a range of exponentially decreasing
star-forming models as well as instantaneous burst models; we have also considered models
with metallicities of solar, 0.4 solar, and 2.5 solar.

The formal best-fit model SED is at a redshift of 2.93, substantially higher than that of 4C\,05.84
itself.  This model has a 0.4-solar-metallicity population with an age of 900 Myr, an 
exponential time constant of 100 Myr, and a reddening $A_V = 0.06$ mag.  If we restrict
ourselves to models with redshifts close to that of the quasar, we get a reasonable fit with
a solar-metallicity model with a redshift of 2.40, an age of 1.02 Gyrs, an exponential time
constant of 200 Myr, and a reddening $A_V = 0.58$ mag. This model fits the $I$-band and 
IRAC 5.8 $\mu$m photometry less well but the IRAC 7.9 $\mu$m photometry slightly better.
Both of these models are shown in Fig.~\ref{4c05sed}. We have no firm grounds 
for choosing one of these SEDs over the other; but, given the uncertainties in the models
and possible star-formation histories, we will accept for the remainder of this paper that the redshift
closer to that of 4C\,05.84 itself is the correct one.  In either case, we are dealing with a massive
galaxy comprising stars that mostly formed $\sim1$ Gyr before the observed epoch.

\begin{figure*}[!bt]
\epsscale{1.0}
\plotone{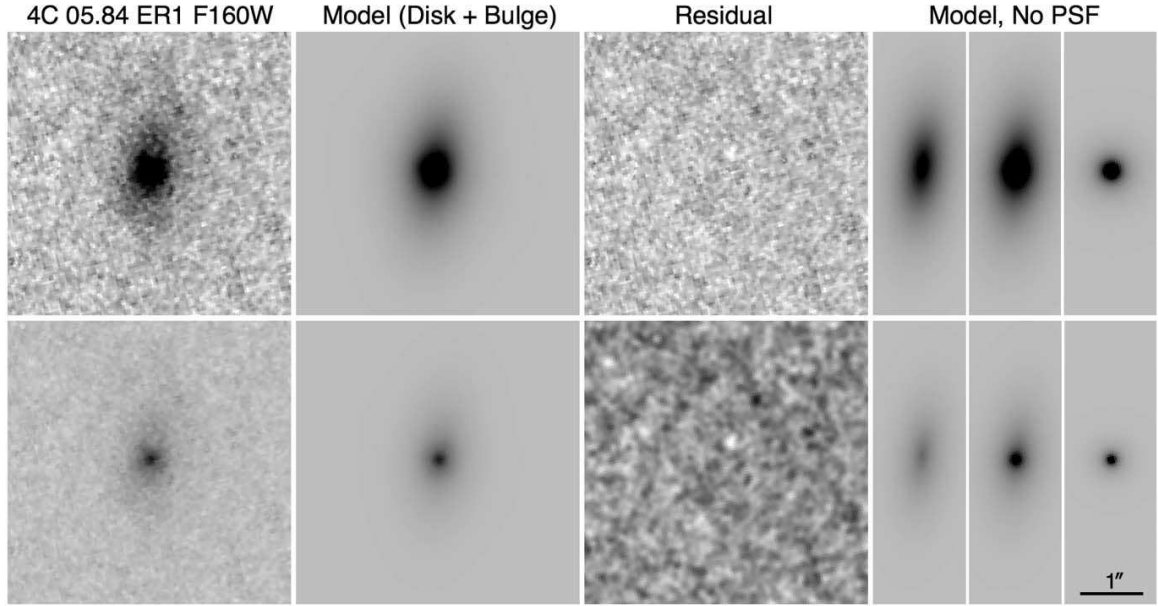}
\caption{The NICMOS2 image of 4C\,05.84\,ER1 in the F160W filter (left panels).
The best-fit {\sc galfit} composite $r^{1/4}$-law + exponential model, 
convolved with the PSF, is shown in the second panel of each 
row, the difference between the observed images and the model in the third
panel, and the model without convolution with the PSF in the last panel. This last
panel is divided into 3 sub-panels:  the middle one of these shows the composite
model, the left one shows the exponential component alone, and the right one
shows the $r^{1/4}$-law subcomponent alone.
The lower panels show lower-contrast images, except for that for the residual image,
which shows a slightly smoothed high-contrast version of the image.}\label{4c05mos}
\end{figure*}

\begin{figure*}[!tb]
\epsscale{1.0}
\plottwo{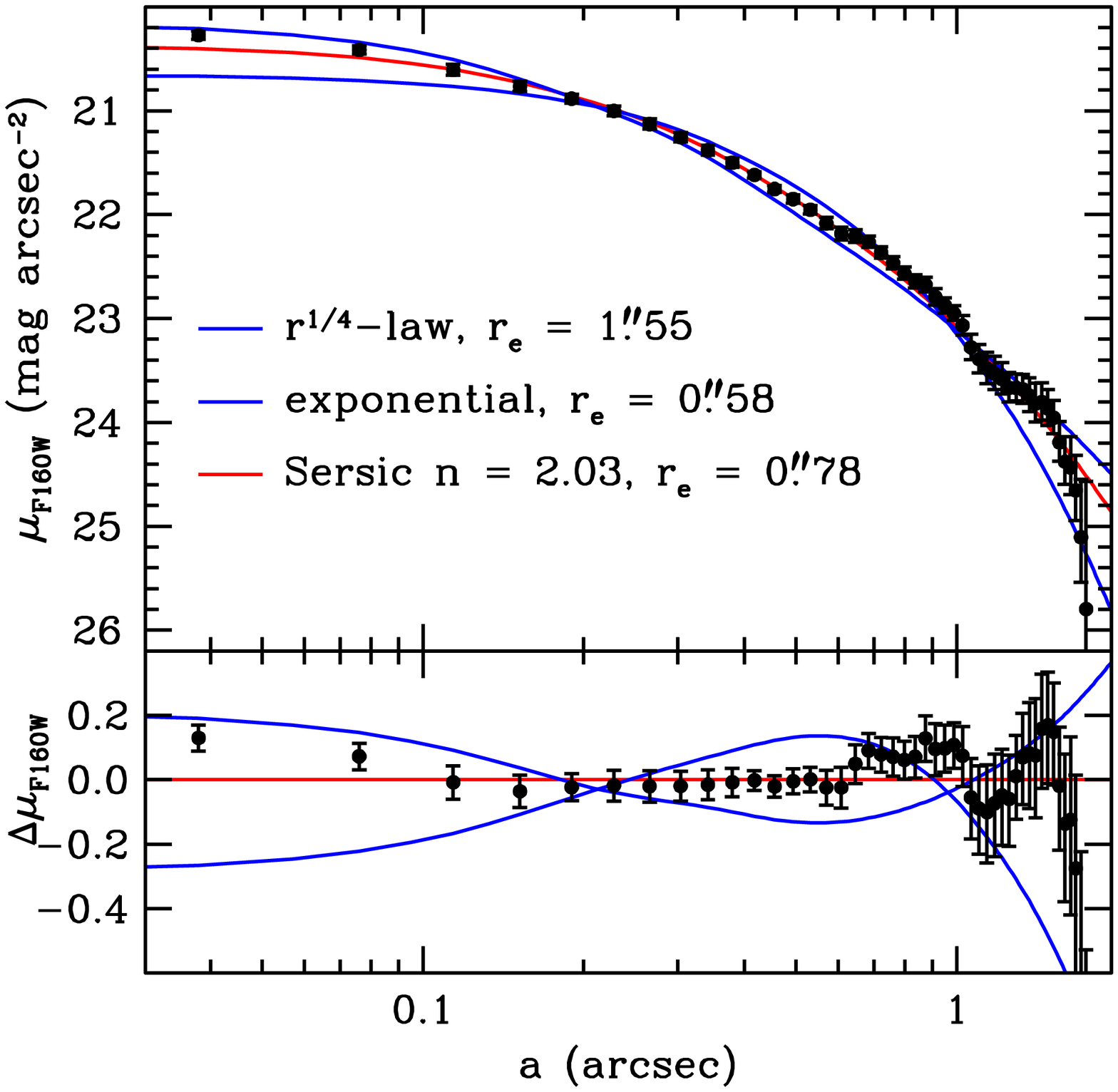}{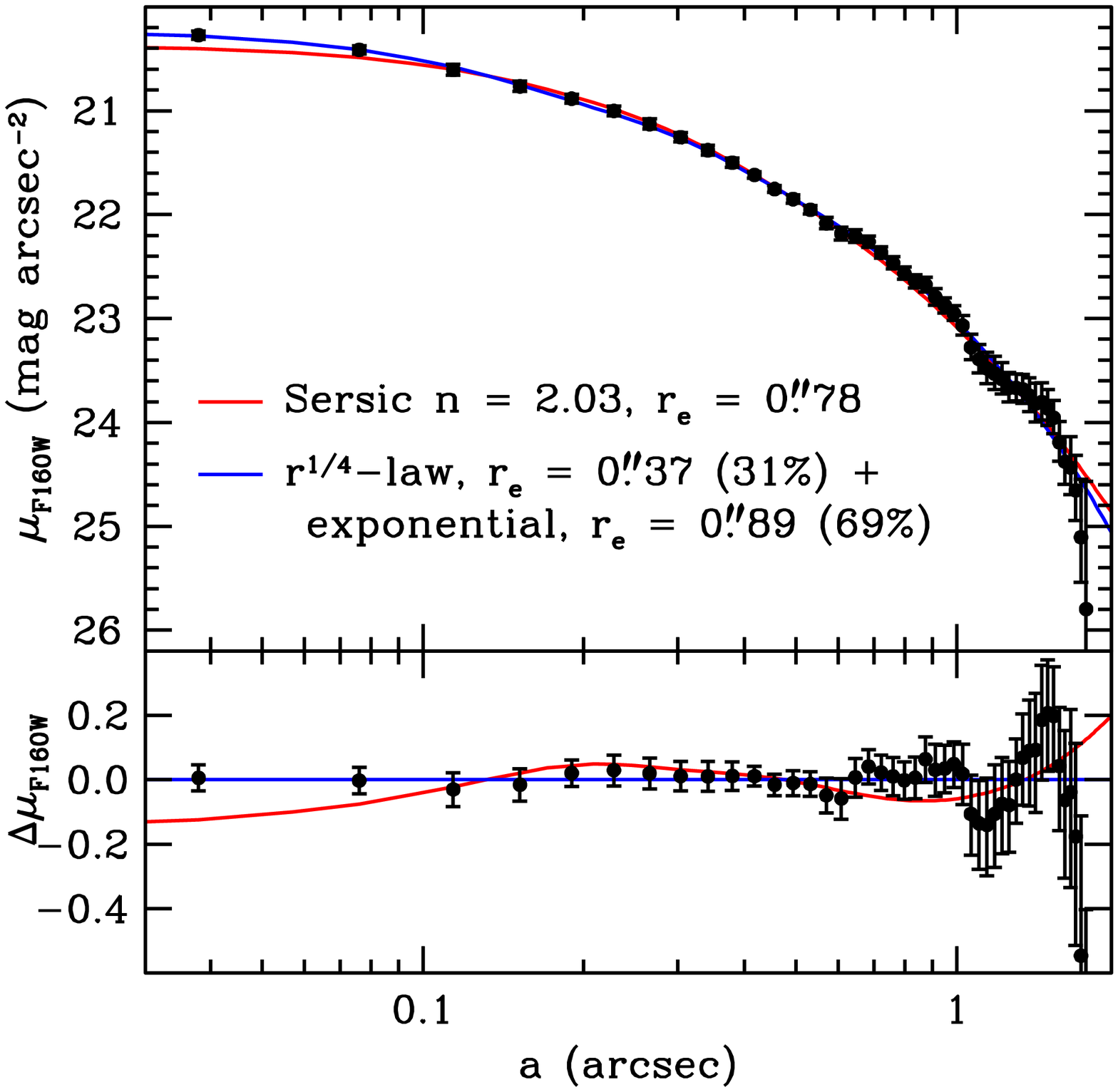}
\caption{Radial-surface-brightness profile of the NICMOS2 F160W image of
\eb.  In the left panel, the best-fit $r^{1/4}$-law, exponential, and S\'{e}rsic profiles
are shown in the same way as in Fig.~\ref{4c23hrsb}.  In the right panel, a composite
model consisting of a small $r^{1/4}$-law bulge, accounting for 31\%\ of the
light, and an exponential disk comprising the remainder, is compared with the
best-fit S\'{e}rsic profile.  In this case, the differentials in the bottom part of the 
panel are made with respect to the composite model.}\label{4c05hrsb}
\end{figure*}

We show our NICMOS2 F160W image of \eb\ in Fig.~\ref{4c05mos}, along with our best-fit
{\sc galfit} model.  We have tried a series of models, including, again, $r^{1/4}$-law,
exponential, and S\'{e}rsic.  Radial-surface-brightness profiles of these are shown in
the left panel of Fig.~\ref{4c05hrsb}.  In this case, even the S\'{e}rsic profile is not a
particularly good fit.  We get a significantly better fit with a two-component model
incorporating a small $r^{1/4}$-law bulge comprising $31 \pm 15$\%\ of the light and an 
exponential disk accounting for the rest.  This model is compared with the best
S\'{e}rsic profile fit in the right panel of Fig.~\ref{4c05hrsb}. For the two-component
model, the disk component has an effective radius $r_e = 0\farcs89 \pm 0\farcs09$,
and the bulge component has $r_e = 0\farcs37 \pm 0\farcs2$.  This best-fit model gives
a total magnitude (on the Vega system) $m_{F160W} = 20.28$.

There is some evidence from our $R$ and $I$-band imaging and our recent 
Keck AO imaging in the $J$ band that the bulge component virtually disappears
at these shorter wavelengths, indicating that the two morphological components
have different stellar populations. Such a result that would not be surprising.  The 
SED shown in Fig.~\ref{4c05sed} would then be the linear combination of the
two SEDs, with the bulge likely being a nearly pure old population and with
younger stars being confined to the disk component.  We stress, however,
that even the disk component must be dominated by old ($\sim$ few hundred
Myr) stars, with little very recent star formation.  We have experimented with
a range of combinations of SEDs at the quasar redshift, but none with simple
star-formation histories (instantaneous bursts or exponentially decaying bursts)
gave a significantly better fit than did the single-population SEDs shown in Fig.~\ref{4c05sed}.
We will explore this possibility in more detail elsewhere.

\section{Discussion\label{discuss}}
Table~\ref{tab1} summarizes the parameters for the two galaxies.
The morphologies of both \ea\ and \eb\ appear to be dominated by disks of 
old stars. However, the disks are quite different in scale.  
 \eb, at least, also appears to have a small bulge
comprising about 1/3 of the total light in the F160W filter ($\sim4800$ \AA,
rest frame). We cannot exclude the possibility that \ea\ also has a weak bulge,
with up to $\sim15$\% of the total light in the F160W filter; indeed, if the slight
apparent difference in morphology between the F160W and F110W images is
real, such a difference would seem to favor this possibility.

But it is the presence of massive, old disks that continues to give the strongest
constraint on formation mechanisms.  Such disks also have been seen at 
redshifts $\sim1.5$, where normal ellipticals with
$r^{1/4}$-law profiles are also found
\citep*{iye03, cim04, yan04, fu05, sto06, mcG07}. It is difficult to 
imagine that these massive disks
could have formed via any process other than the dissipative collapse of a large
cloud of gas.  Such disks are also unlikely to have survived major merging events,
although the bulge component in \eb\ may testify to either some level of minor
merging activity or bulge building via disk instabilities. 

For galaxies at $z\sim2.5$, the evidence for a dominant old stellar population
depends on the inflection in the SED shortward of the $H$ band, and establishing
this inflection with optical/near-IR photometry depends on the relatively short baseline 
from the $H$ to the $K$
band.  Furthermore, at the present epoch, essentially all strongly disk-dominated
galaxies show evidence for continued star formation.  It is therefore not too
surprising that claims of passive disks at high redshift should be doubted 
\citep[\eg][]{pie05}.  However, as Fig.~\ref{4c05sed} shows, {\it Spitzer} IRAC data
is entirely consistent with the SED of a moderately old stellar population, and
no plausible SED incorporating very recent star formation combined with dust would fit
the observed photometry. We have recently also obtained IRAC
imaging of the field of 4C\,23.56, and our analysis of these data
shows that the IRAC photometry falls squarely on our best-fit 
solar-metallicity BC03 model determined from
the optical/near-IR photometry alone: an instantaneous burst with an age of
2.6 Gyr and an extinction $A_V=0.16$ mag \citep{sto07}. Using the more recent
preliminary CB07 models, with their improved treatment of AGB stars,
we obtain a stellar population age of 2.8 Gyr with $A_V=0$.  Again, no plausible model
with significant star formation and reddening would fit these data.
\begin{deluxetable}{l c c c c}
\tablewidth{0pt}
\tablecaption{Model Parameters for \ea\ and \eb}
\tablehead{
\colhead{Galaxy} & \colhead{Filter} & \colhead{S\'{e}rsic $n$} & \colhead{$r_e$} & \colhead{$r_e$}\\
  & & & (\arcsec) & (kpc)
}
\startdata
4C\,23.56\,ER1 & F110W & $1.03\pm0.10$ & $0\farcs28\pm0\farcs02$ & $2.2\pm0.2$ \\
4C\,23.56\,ER1 & F160W & $1.52\pm0.06$ & $0\farcs24\pm0\farcs01$ & $1.9\pm0.1$ \\
                            &               & 1.00\tablenotemark{a} & $0\farcs89\pm0\farcs09$ & $7.1\pm0.8$ \\
\raisebox{1.5ex}[0pt]{4C\,05.84\,ER1} & \raisebox{1.5ex}[0pt]{F160W} & 4.00\tablenotemark{a} & $0\farcs37\pm0.20$ & $3.0\pm1.6$ \\
\enddata
\tablenotetext{a}{The S\'{e}rsic indices for the two model components for 4C\,05.84\,ER1 have been fixed at these values, 
which correspond to exponential and $r^{1/4}$-law profiles, respectively.}
\label{tab1}
\end{deluxetable}

Masses for these galaxies can be estimated from the model fits.  Assuming
solar metallicities and a \citet{cha03} initial mass function, we obtain a mass
of $3.9\times10^{11} M_{\odot}$ for \ea\ and $3.3\times10^{11} M_{\odot}$
for \eb (assuming the model at $z=2.4$ with $A_V=0.58$).

While the stellar-population age of \eb\ indicates that the last
major star-formation episode occurred at $z\sim3.7$, when the universe was
$\sim1.8$ Gyr old, \ea\ has a stellar-population age that is formally slightly greater
than the age of the universe at $z=2.483$.  Clearly, the likely errors in the age
determination and the usual caveats regarding the age-metallicity degeneracy
mitigate any implied paradox.  Nevertheless, this massive galaxy must have formed
at a very high redshift.  Models with [Fe/H] $= +0.4$ give an age of 1.9 Gyr,
but with a significantly worse fit.

It therefore seems likely that galaxy formation models will have to allow for the
presence of early-forming massive disks.  This means that, at least in some
dense regions, it has been possible to form $\sim3\times10^{11}$ $M_{\odot}$
of stars within a relatively short time via dissipative collapse and without the aid of
major mergers.  While our selection criteria have ensured that the galaxies we
have discussed here comprise essentially pure old stellar populations, they
may well be representative of many massive galaxies at high redshift, most
of which would not be in our sample if they retained even tiny amounts of residual star
formation or if they had had any significant star formation within a few hundred Myr prior to
the epoch at which we observe them.

\eb\ has a luminosity and an effective radius that are similar to those of many
local galaxies.  Our best-fitting S\'{e}rsic model has $r_e=6.3$ kpc.  For comparison,
for galaxies of similar mass from the Sloan survey with S\'{e}rsic $n<2.5$, \citet{she03} find
$r_e=7.2^{+2.9}_{-2.1}$ kpc.  This galaxy could become, with passive evolution and perhaps 
a few minor mergers to increase the bulge-to-disk ratio somewhat, a typical S0 galaxy at
the present epoch.  On the other hand, we do not see galaxies like \ea\ at the present
epoch.  By the prescription of \citet{she03}, a low-S\'{e}rsic-index galaxy with the mass
of \ea\ would have $r_e=7.6^{+3.1}_{-2.2}$ kpc, but \ea\ actually has $r_e=1.9\pm0.1$
kpc. This means that the stellar mass surface density is much higher than for
local galaxies, a result that has also been found for other samples of distant red 
galaxies \citep[\eg][]{tru06,tof07}.  It would seem that the only likely path for such galaxies
to evolve to objects consistent with the local population of galaxies is through
dissipationless mergers.

There is recent evidence that the most massive galaxies in the local
universe are likely the result of dry mergers of galaxies with stars that are already 
old and with very little gas \citep[e.g.,][]{ber07}.  With the constraint that these
merging components must themselves mostly be fairly massive (to avoid a large
dispersion and flattening in the observed color---magnitude relation for 
present-day massive galaxies, \eg\ \citealt*{bow98}), it seems possible that 
these early massive disks may well be among the sources for the old stars that 
today are found in the most massive elliptical galaxies.

\acknowledgments
We thank S. Charlot and G. Bruzual for providing us with preliminary versions of
their new spectral synthesis models prior to publication.  We also thank the anonymous
referee for a detailed reading of the paper and a number of specific suggestions that helped
us improve it.
Support for {\it HST} program no.~10418 was provided by NASA through a grant from 
the Space Telescope Science Institute, which is operated by the Association 
of Universities for Research in Astronomy, Inc., under NASA contract NAS 5-26555.
This research has also been partially supported by NSF grant AST03-07335. 
It made use of the NASA/IPAC Extragalactic Database (NED) 
which is operated by the Jet Propulsion Laboratory, California Institute of 
Technology, under contract with the National Aeronautics and Space 
Administration.


\begin{thebibliography}{}
\bibitem[Abraham \etal(2007)]{abr07} Abraham, R. G. \etal\ 2007, \apj, submitted 
   [astro-ph/0701779]
\bibitem[Bergeron \& Dickinson(2003)]{ber03} Bergeron, L. E., \& Dickinson, M. 2003,
   NICMOS Instrument Science Report 2003-010 
   [{\tt http://www.stsci.edu/hst/nicmos/ documents/isrs/isr\_2003-10.pdf}]
\bibitem[Bernardi \etal(2007)]{ber07} Bernardi, M., Hyde, J. B., Sheth, R. K., Miller, C. J.,
   \& Nichol, R. C. 2007, \aj, 133, 1741
\bibitem[Bower \etal(1998)Bower, Kodama, \& Terlevich]{bow98} Bower, G., Kodama, T., 
   \& Terlevich, A. 1998, \mnras, 299, 1193
\bibitem[Bruzual \& Charlot(2003)]{bru03} Bruzual, G., \& Charlot, S. 2003, \mnras, 344, 1000
\bibitem[Calzetti \etal(2000)]{cal00} Calzetti, D., Armus, L., Bohlin, R. C., Kinney, A. L.,
   Kornneef, J., \& Storchi-Bergmann, T. 2000, \apj, 533, 682
\bibitem[Cimatti \etal(2004)]{cim04} Cimatti, A., \etal\ 2004, Nature, 430, 184
\bibitem[Chabrier(2003)]{cha03} Chabrier, G. 2003, \pasp, 115, 763
\bibitem[Charlot \& Bruzual(2007)]{cha07} Charlot, S., \& Bruzual, G. 2007, in preparation
\bibitem[Daddi \etal(2005)]{dad05} Daddi, E., \etal\ 2005, \apj, 626, 680
\bibitem[de Propris \etal(2007)]{deP07} De Propris, R., Stanford, S. A., Eisenhardt, P. R.,
   Holden, B., \& Rosati, P. 2007, \aj, in press [astro-ph/0702050]
\bibitem[Fruchter \& Hook(2002)]{fru02} Fruchter, A. S., \& Hook, R. N. 2002, \pasp, 114, 144
\bibitem[Fu \etal(2005)Fu, Stockton, \& Liu]{fu05} Fu, H., Stockton, A., \& Liu, M. 
   2005, \apj, 632, 831
\bibitem[Hawarden \etal(2001)]{haw01} Hawarden, T. G., Leggett, S. K.,
   Letawsky, M. B., Ballantyne, D. R., \& Casali, M. M. 2001, \mnras, 325, 563
\bibitem[Iye \etal(2004)]{iye04} Iye, M. \etal\ 2004, \pasj, 56, 381
\bibitem[Iye \etal(2003)]{iye03} Iye, M. \etal\ 2003, \apj, 590, 770
\bibitem[Knopp \& Chambers(1997)]{kno97} Knopp, G. P., \& Chambers, K. C. 1996, \apjs, 109, 367
\bibitem[Kobayashi \etal(2000)]{kob00} Kobayashi, N., \etal\ 2000, in Proc. SPIE 4008: Optical and 
IR Telescope Instrumentation and Detectors, ed. M. Iye \& A. F. Moorwood, p. 1056  
\bibitem[Kriek \etal(2006)]{kri06} Kriek, M., \etal\ 2006, \apj, 649, L71
\bibitem[Labb\'{e} \etal(2005)]{lab05} Labb\'{e}, I. \etal\ 2005, \apj, 624, L81
\bibitem[Landolt(1992)]{lan92} Landolt, A. U. 1992, \aj, 104, 340
\bibitem[Leggett \etal(2006)]{leg06} Leggett, S. K. \etal\ 2006, \mnras, 373, 781
\bibitem[McCarthy \etal(2004)]{mcC04} McCarthy, P. \etal\ 2004, \apj, 614, L9
\bibitem[McGrath \etal(2007)]{mcG07} McGrath, E. J., Stockton, A., Canalzio, G.,
   Iye, M., \& Maihara, T. 2007, \apj, submitted
\bibitem[Maraston(2005)]{mar05} Maraston, C. 2005, \mnras, 362, 799
\bibitem[Marigo \& Girardi(2007)]{mar07} Marigo, P., \& Girardi, L. 2007, \aap, 469, 239
\bibitem[Motohara \etal(2002)]{mot02} Motohara, K., \etal\ 2002, \pasj, 54, 315  
\bibitem[Nelan \etal(2005)]{nel05} Nelan, J. E., Smith, R. J., Hudson, M. J., Wegner, G. A.,
   Lucey, J. R., Moore, S. A. W., Quinney, S. J., \& Suntzeff, N. B. 2005, \apj, 632, 137
\bibitem[Papovich \etal(2006)]{pap06} Papovich, C. \etal\ 2006, \apj, 640, 92
\bibitem[Peebles(2002)]{pee02} Peebles, P. J. E. 2002, ASP Conf. Ser., 283, 351
\bibitem[Peng \etal(2002)]{pen02} Peng, C. Y., Ho, L. C., Impey, C. D., \& Rix, H.-W. 2002,
   \aj, 124, 266
\bibitem[Pierini et al.(2005)]{pie05} Pierini, D., Maraston, C., Gordon, K.~D., \& Witt, A.~N. 2005, 
   \mnras, 363, 131
\bibitem[Reddy \etal(2006)]{red06} Reddy, N. A., Steidel, C. C., Fadda, D., Yan, L., Pettini, M.,
Shapley, A. E., Erb, D. K., \& Adelberger, K. L. 2006, \apj, 644, 792
\bibitem[Renzini(2006)]{ren06} Renzini, A. 2006, \araa, 44, 141
\bibitem[Scarlata \etal(2007)]{sca07} Scarlata, C. \etal\ 2007, \apj, in press [astro-ph/0701746]
\bibitem[Schlegel, Finkbeiner, \& Davis (1998)]{sch98} Schlegel, D. J., Finkbeiner, D. P., \& 
   Davis, M. 1998, \apj, 500, 525
\bibitem[Sheinis \etal(2002)]{she02} Sheinis, A. I., Bolte, M., Epps, H. W., Kibrick, R. I.,
   Miller, J. S., Radovan, M. V., Bigelow, B. C., \& Sutin, B. M. 2002, \pasp, 114, 851
\bibitem[Shen \etal(2003)]{she03} Shen, S., Mo, H. J., White, S. D. M., Blanton, M. R., Kauffmann,
   G., Voges, W., Brinkmann, J., \& Csabai, I. 2003, \mnras, 343, 978
\bibitem[Stockton \etal(2004)Stockton, Canalizo, \& Maihara]{sto04} Stockton, A., Canalizo, G., \&
   Maihara, T. 2004, \apj, 605, 37
\bibitem[Stockton \& McGrath(2007)]{sto07} Stockton, A., \& McGrath, E. 2007, ASP Conf Ser,
   in press [astro-ph/0702130]
\bibitem[Stockton \etal(2006)Stockton, McGrath, \& Canalizo]{sto06} Stockton, A., 
   McGrath, E., \& Canalizo, G. 2006, \apj, 650, 706
\bibitem[Takami \etal(2004)]{tak04} Takami, H., \etal\ 2004, \pasj, 56, 225
\bibitem[Thomas \etal(2005)]{tho05} Thomas, D., Maraston, C., Bender, R., \& 
   Mendez de Oliveira, C. 2005, \apj, 621, 673
\bibitem[Toft \etal(2007)]{tof07} Toft, S., \etal\ 2007, \apj, in press [arXiv:0707.4484]
\bibitem[Trujillo \etal(2006)]{tru06} Trujillo, I. \etal\ 2006, \apj, 650, 18
\bibitem[van Dokkum(2001)]{vanD01} van Dokkum, P. G. 2001, \pasp, 113, 1420
\bibitem[van Dokkum(2004)]{vanD04} van Dokkum, P. G., \etal\ 2004, \apj, 611, 703
\bibitem[Yan \& Thompson(2003)]{yan03} Yan, L., \& Thompson, D. 2003, \apj, 586, 765
\bibitem[Yan \etal(2004)Yan, Thompson, \& Soifer]{yan04} Yan, L., Thompson, D., \& Soifer, B. T. 2004, 
   \aj, 127, 1274
\bibitem[Zirm \etal(2007)]{zir07} Zirm, A. W., \etal\ 2007, \apj, 656, 66
\end{thebibliography}
\end{document}